\documentclass[12pt,preprint]{aastex}
 
\shorttitle{He I $\lambda$10830 in Young Stars}
\shortauthors{Edwards et al.}
 
\slugcomment{}

\shorttitle{He I $\lambda$10830 in Young Stars}
\shortauthors{Edwards et al.}
\slugcomment{Accepted ApJ Letters}
\begin{document}

\title{He I $\lambda$10830 as a Probe of Winds in Accreting Young
Stars }

\author{Suzan Edwards \altaffilmark{1,2,3}, William
Fischer\altaffilmark{4} , John Kwan \altaffilmark{4},
Lynne Hillenbrand \altaffilmark{2,5},
and A.K. Dupree \altaffilmark{1,6}}

\altaffiltext{1}{Guest Observer, Gemini/Keck-NIRSPEC program}
\altaffiltext{2}{Visiting Astronomer, Keck Observatory}
\altaffiltext{3}{Five College Astronomy, Smith College,
 Northampton, MA 10163, sedwards@smith.edu}
\altaffiltext{4}{Five College Astronomy, University of
 Massachusetts, Amherst, MA, 01003
 kwan@nova.astro.umass.edu, wfischer@nova.astro.umass.edu}
\altaffiltext{5} {Dept. of Astronomy, California
Institute of Technology, Pasadena, CA 91125,
lah@astro.caltech.edu}
\altaffiltext{6} {Harvard-Smithsonian Center for
Astrophysics, Cambridge, MA 01238, dupree@cfa.harvard.edu}

\begin{abstract}

He I $\lambda$10830 profiles acquired with Keck's NIRSPEC for 6
young low mass stars with high disk accretion rates
(AS 353A, DG Tau, DL Tau, DR Tau, HL Tau and SVS 13) provide new
insight into accretion-driven winds. In 4 stars the profiles
have the signature of resonance scattering, and possess a deep
and broad blueshifted absorption that penetrates more than 50\% into
the 1 $\micron$ continuum over a continuous range of velocities
from near the stellar rest velocity to the terminal velocity of
the wind, unlike inner wind signatures seen in other spectral
features. This deep and broad absorption provides the first
observational tracer of the acceleration region of the inner
wind and suggests that this acceleration region is situated
such that it occults a significant portion of the stellar
disk. The remaining 2 stars also have blue absorption
extending below the continuum although here the profiles are
dominated by emission, requiring an additional source of
helium excitation beyond resonant scattering. This is likely
the same process that produces the emission profiles seen at
He I $\lambda$5876.

\end{abstract}


\section{Introduction}

Outflows from young stars undergoing disk accretion are suspected
to play a vital role in the process of star formation and in the
evolution of protoplanetary disks. Although their ubiquity among
accreting YSOs is well documented, and there is some evidence
for a rough proportion of ${\dot M}_{wind}/{\dot M}_{acc} \sim
0.1$ \citep{kon00}, the origin of the outflows
remains a mystery. Most outflow models tap
magneto-centrifugal ejection as the heart of the launching
mechanism, where rotating fields in the disk fling material
along inclined field lines. A distinguishing characteristic
among disk wind models is the location in the system where
mass loading onto field lines occurs -- either from the
inner disk over a range of radii \citep{kon00} or from the
radius at which the stellar magnetosphere truncates the
disk, lifting most of the accreting material toward the star
into magnetic funnel flows and ejecting the rest along
opened stellar field lines emerging from this point
\citep{shu94}. Others have explored the possibility that
accretion-driven winds emerge via magnetohydrodynamic
acceleration from the star rather than, or in addition to,
the disk \citep{hir97,kwa88}. Each approach has different
implications for angular momentum evolution in accretion
disk systems, both for the disk, where the dominant mode of
angular momentum transport remains unknown, and for the
accreting star, where the means of stellar spin-down in the
face of accumulation of high angular momentum material is
uncertain.

Observational probes of winds in accreting systems have not yet
provided decisive evidence for where the wind originates. Either
outflowing gas is observed far from the launch site, i.e. in
collimated jets and molecular outflows, or else inner winds
close to the star produce kinematic signatures that are
entangled with emission deriving from other phenomena such as
funnel flows and accretion shocks associated with
magnetospheric accretion. Among the classical T Tauri stars (cTTS), the
traditional indicator of an inner wind is blueshifted absorption
superposed on strong and broad emission lines, as seen in H$\alpha$, Na D,
Ca II H\&K, and MgII h\&k \citep{naj00}. The line
emission is attributed primarily to infall in magnetic funnel flows
\citep{lee94, muz01} while the accompanying blueshifted
absorption signifies the presence of a high velocity wind
close to the star \citep{cal97, ard02}. However little
progress has been made in understanding the properties of
this inner wind or in establishing whether it originates
from a range of radii in the inner disk, the disk truncation
radius, or the star.

A recent analysis of profiles of He I $\lambda$5876 and
$\lambda$6678 in a sample of 31 accreting cTTS revealed a new
means of diagnosing the inner wind \citep{ber01}. The helium
lines were selected for study because of their high excitation
potential, restricting line formation to a region either of high
temperature or close proximity to a source of ionizing
radiation. In spite of these restrictions, the helium
emission lines were found to have a composite origin,
including contributions from a wind, from the funnel flow,
and from an accretion shock.  The wind component,
characterized by broad blueshifted emission (centroid
velocities exceeding $-$30 km~s$^{-1}$) and extended blue
wings (maximum velocities exceeding $-$200 km~s$^{-1}$), was
most prominent among stars with the highest disk accretion
rates. Curiously, the high accretion rate stars with a
strong hot helium wind component were also found to have
anomalously weak or absent signatures of the narrow emission
component formed in the accretion shock where the
magnetosphere directs accreting material from the disk. We
interpreted these findings as indicating that a hot
accretion driven wind originates in the vicinity of the
stellar corona, and that when the hot wind is present, the
conditions of magnetospheric accretion and its accompanying
accretion shock may be modified.

This intriguing result led us to pursue an investigation of an
additional line in the He I triplet series, $\lambda$10830,
immediately following $\lambda$5876 in a recombination-cascade
sequence. The lower $2s^3S^{\circ}$ level of He I $\lambda$10830 is
metastable, resulting in conditions ripe for resonance
scattering, which provides an opportunity to search for
outflowing gas in absorption. The sensitivity of $\lambda$10830
absorption to the dynamics of winds in cool stars
is illustrated in Dupree et al. (1992). This feature is
in a spectral region that was until recently not readily accessible to
either CCD or infrared imaging detectors. An early study by
Ulrich and Wood in 1981, using pre-ccd technology, detected this
feature in a few T Tauri stars, but the data were not of
sufficient quality to interpret the line profile. The next
published work featuring $\lambda$10830 in a T Tauri star is for
DG Tau {\citep{tak02}, showing broad emission and a blueshifted
absorption feature, confirming the prediction of BEK that the inner
wind is traced by helium.

Using Keck II's NIRSPEC, we are undertaking a
census of He I $\lambda$10830 in accreting young stars to evaluate
the constraints that helium places on inner winds.
In this Letter, we present He I $\lambda$10830 profiles for 6
low mass YSOs with higher than average disk accretion and mass
outflow rates to demonstrate that this feature offers new insight
into mass loss from young stars with accretion disks.

\section{Observations}

The 6 YSOs we have selected to demonstrate the diagnostic
potential of He I $\lambda$10830 are listed in
Table~\ref{tbl-1}, along with some outflow and accretion properties
and observed characteristics of their $\lambda$10830
profiles. These stars are among the cTTS with the highest
disk accretion rates, as determined from their large optical
veiling. Most accretion rates from Table 1 are from
Hartigan, Edwards and Ghandour (1995, hereafter HEG), and
are about an order of magnitude higher than the average
found for cTTS in that study, ${\dot M}_{acc}$ $\sim
10^{-7} M_{\odot} yr^{-1}$. A more recent analysis
of cTTS disk accretion rates \citep{gul98} puts the
average rate at ${\dot M}_{acc}$ $\sim 10^{-8} M_{\odot}
yr^{-1}$, but these high accretion rate stars were
not included in that analysis. As also noted in Table 1, our
sample stars all have collimated jets traced by high
velocity forbidden line emission, and all have inner winds
traced by blueshifted absorption at $H\alpha$. Five (AS
353A, DG Tau, DR Tau, HL Tau and SVS 13) are extreme Class
II sources \citep{lad87}, although two of these (HL
Tau and SVS 13) are sometimes included in compilations of
Class I sources, presumably due to their large extinction
and well developed jets and molecular outflows. The sixth
(DL Tau) is a more typical cTTS.  In addition, we have also
observed the "weak" T Tauri star (wTTS) V827 Tau, a
non-accreting star of comparable mass and age to the 6 cTTS.

Spectra were acquired with NIRSPEC on Keck II in its echelle
mode with the N1 filter (Y band), providing wavelength coverage from
0.95-1.12 $\mu m$ at R = 25,000. Spectra of DG Tau and HL
Tau were taken on 22 November 2001 by the Gemini-Keck queue
observers T. Geballe and M. Takamiya. The remaining spectra were taken
during a 3 night run in November 2002 by Hillenbrand and Edwards.
Spectra were processed in IRAF, and were spatially rectified with
a reduction script written and provided by M. Takamiya.
Spectra were corrected for telluric emission by subtracting pairs of
spectra taken at two positions along the 12" slit, and for
telluric absorption by dividing each spectrum by an early-type
standard taken at similar airmass. During the 2002 run spectra
were acquired without the image de-rotator, due to equipment
malfunction. Although this prevents studying spatially
extended emission, as reported for DG Tau \citep{tak02}, it allows
extraction of a spectrum from the central point source, where
most of the helium emission and all of the helium absorption arises.

\section{Discussion}

The He I $\lambda$10830 profiles for 6 high accretion rate cTTS
are shown in Figures 1 and 2. The
$\lambda$10830 profiles all show unequivocal evidence for mass
loss with blueshifted absorption penetrating below the
continuum level, with depths ranging from 20\% (DG Tau) to 90\% (DR Tau)
and with maximum velocites ranging from $-$200 km~s$^{-1}$ (HL
Tau) to $-$450 km~s$^{-1}$(DG Tau). In Figure 1 we compare them to
non-simultaneous profiles for He I $\lambda$5876 from BEK for
the 4 stars in common with that study and in Figure 2 we compare
them to non-simultaneous H$\alpha$ profiles from BEK or Hillenbrand and
White (in preparation).  While blueshifted absorption
is not seen in the $\lambda$5876 transition (discussed below),
it is present in traditional indicators of the inner wind,
such as the H$\alpha$ profiles shown here. However what
distinguishes He I $\lambda$10830 is both the prominence and
depth of its blueshifted absorption, offering the promise of a
more direct probe of the inner wind. At H$\alpha$ all profiles
are overwhelmingly dominated by broad and strong emission, while
for $\lambda$10830 4 of the 6 stars are either predominantly in
absorption (DR Tau and SVS 13) or else resemble classic P Cygni
profiles, with comparable levels of emission and absorption (HL
Tau and AS353A). Moreover at $\lambda$10830 the blue absorption
is far deeper and broader than at H$\alpha$, penetrating the
continuum in all 6 stars. The most significant aspect of the
blueshifted absorption at helium is that in 4 stars (DR Tau, HL
Tau, SVS 13 and AS 353A) it penetrates to depths of more than 50\% into
the continuum over a continuous and broad velocity range
extending from near the stellar rest velocity to the terminal
wind velocity, as inferred from velocities characterizing their
spatially extended jets (see Table 1).

The uniqueness of He I $\lambda$10830 as a probe of the inner wind
derives from the metastability of its lower level ($2s^3S^{\circ}$),
which, although energetically far above (20 eV) the singlet ground
state, is radiatively isolated from it. Whether this metastable
level is populated by recombination and cascade or by
collisional excitation from the ground state, it will become
significantly populated relative to other excited levels owing
to its weak de-excitation rate via collisions to singlet states,
making it an ideal candidate to form an absorption line. This
absorption is essentially a resonant-scattering process since
the $\lambda$10830 transition is the only permitted radiative
transition from its upper state to a lower state and the
electron density is unlikely to be so high as to cause
collisional excitation or de-excitation. The absence of any
absorption or emission at $\lambda$10830 in the non-accreting
wTTS V827 Tau (not shown), which like all wTTS has strong
coronal x-rays \citep{fei02}, clarifies that the presence of
an inner T Tauri wind traced via resonant scattering from
the helium metastable level requires the additional
presence of disk accretion and its associated energetic phenomena.

New insight regarding the inner wind in cTTS can be made from visual
inspection of the helium profiles for those 4 stars
where resonance scattering
is the dominant formation mechanism. Their deep and broad
absorption troughs reveal new information on the kinematics of the
inner wind, probing its acceleration region in an accreting star for
the first time, and on the geometry of the inner wind,
suggesting that the acceleration region is likely close to the
star and occults a significant portion of it.

The insight into the kinematics of the wind is provided by
the large and continuous span of velocities over which the
blueshifted absorption is observed. This suggests that we
are probing the acceleration region of the inner wind, as
the absorption extends from velocities at rest relative to
the star continuously through to velocities comparable to or
often larger than the terminal wind velocities inferred from
spatially extended jets. There is also a suggestion of
turbulence in the inner wind, since in 2 of the stars the
blueshifted absorption trough extends to velocities as red
as 50 km~s$^{-1}$.

The insight into the geometry of the inner wind comes from
the depth of the absorption into the $1~\mu m$ continuum
over such a wide range of velocities, from rest velocity to
the terminal velocity of the wind. In fact, the actual
depletion of blueshifted continuum photons will exceed what
is observed, as some photons scattered into the line of
sight will fall within the velocity range of the blue
absorption, making it appear shallower and less extended. At
$1~\mu m$ the continuum in a heavily veiled star will have a
larger fraction contributed by the stellar photosphere than
at other wavelengths, because the spectral energy
distributions of the excess emission arising from both the
accretion shock on the stellar surface \citep{gul98} and the
inner disk \citep{muz03} are rapidly decreasing toward this
wavelength while the stellar spectral energy distribution is
peaking. For example, at $1~\mu m$ veiling from an accretion
shock at T$\sim$8000K will be reduced by a factor of 5
compared to the V band and veiling from the inner disk with
T$\sim$1400K will be reduced by a factor of 4 compared to
the K band.

We have determined the veiling at $1~\mu m$ for our high
accretion rate stars by comparing the depth of photospheric
features in the echelle order centered at 1.075$\mu m$ with
the wTTS V827 Tau. Our veiling estimates, listed in Table 1,
indicate that the photosphere contributes between 30\% to
50\% of the $1~\mu m$ continuum in stars known for much
higher veilings in the optical and at 2$\mu m$ \citep
{har95,muz03}. (Photospheric lines could not be detected in
SVS 13 because of obscuration by emission lines, however we
estimate the veiling is $\sim 2$ based on the equivalent
width of these lines compared to emission features in the
other stars.) With the clarification that a significant
fraction of the continuum arises from the star, including
both the stellar photosphere and the accretion shock on the
stellar surface, then the implication of the observed depth
and velocity range of the absorption in the 4 stars where
$\lambda$10830 is predominantly formed by resonance
scattering is that the full span of the wind acceleration
region lies along the line of sight to the stellar disk.

An unusual feature of the $\lambda$10830 profiles shown
here is that 2 of the 4 stars with resonance scattering
profiles have significant net negative equivalent widths
(see Table 1). This is unexpected in the standard formation
scenario for a P-Cygni line by resonant scattering, which
yields a net zero equivalent width in the absence of stellar
occultation, indicating that in these two stars far fewer
photons are being scattered into the line of sight than out
of it. In the most extreme case, DR Tau, the net equivalent
width has a value of -10.2 \AA~ or, equivalently in velocity
units, of 283 km~s$^{-1}$. We can demonstrate analytically
that this unusual behavior can be accounted for in a resonant scattering
scenario if occultation of scattered photons is provided by
both the star and the accretion disk. For example, if the
disk were to extend into the stellar surface, thereby
screening photons scattered from receding material, then in
a pole-on system with a spherical isotropically scattering
shell at R = 2 R$_\star$, the photons scattered into the
line of sight amount to only 40\% of the continuum photons
scattered out of the line of sight. This fraction increases
to 76\% as the inclination changes from 0$^{\circ}$ to
90$^{\circ}$. The corresponding net negative equivalent
widths would thus be characterized by 60\% to 24\% removal
of the continuum, values consistent with the large negative
values found here for DR Tau and SVS 13. Moving the scattering
surface to different radii does not significantly alter this result,
although if the disks of these
2 stars are truncated at a radius outside the scattering surface
then accounting for their large negative equivalent widths
will be a challenge.

Additional perspective on the inner wind comes from comparing
optical He I profiles to those of $\lambda$10830. In Figure
1 we have compared (non-simultaneous) $\lambda$10830 and
$\lambda$5876 profiles for the 4 stars also in the BEK study.
For these 4 high accretion rate stars BEK concluded
that their $\lambda$5876 emission primarily originates in a
wind rather than in funnel flows or accretion shocks. This
conclusion was based on the large blueshifted centroids
and/or the large blue wing velocities found in the broad
component of the $\lambda$5876 emission coupled with no
conspicuous narrow component coming from the accretion
shock. This interpretation is strengthened by the
unequivocal wind signatures found at $\lambda$10830 in the
cTTS studied here, where the blue wings are now in
absorption rather than emission. The absence of blue
absorption at $\lambda$5876 is readily understood as due to
the optical thinness of this line in the wind, given the
strong radiative decay rate of the lower level of the
$\lambda$5876 transition.

The strength of the emission at $\lambda$5876 also
suggests that the helium wind can have a significant
optical depth at $\lambda$10830. As this optical depth increases,
helium emission will be enhanced over the resonant
scattering process, either via recombination/cascade
and/or collisional excitation. This effect could then
account for the 2 stars investigated here, DG Tau and DL
Tau, which have large positive equivalent widths at
$\lambda$10830. They could still have an underlying trough-like
absorption from resonance scattering, extending from the
stellar rest velocity to the wind terminal velocity as seen
in the other 4 stars, which would be filled in
by the additional emission processes in the wind.
Since both resonant scattering and line emission
contribute to $\lambda$10830 formation, while the primary
contributor to $\lambda$5876 is line emission, we anticipate
that simultaneous observations of both He I lines, together
with theoretical modelling, may yield the velocity, density,
and temperature structures of the inner wind.

In summary, we conclude that the He I $\lambda$10830 line offers
a unique new probe of inner winds in accretion disk systems,
revealing the wind acceleration region and providing constraints
on the geometry of this region. Most importantly, the deep and
broad absorption seen in 4 stars seems to require that the full
span of the wind acceleration region is seen in projection
against the stellar disk. We find it more straightforward to
interpret this finding in a scenario where the wind arises from
stellar coronal regions, since it would be difficult for wind
material arising from a disk to absorb the
radiation from the stellar disk at all velocities,
from rest to the terminal wind speed. If this is the
case, it would not be a normal "stellar wind", since this
phenomenon is strongest in stars with the highest disk accretion
rates. We anticipate a fuller understanding of the physical
properties of the inner wind, its geometry and its launch
distance from the star when the behavior of this line has been
examined for a large sample of accreting young stars covering a
wide range of disk accretion rates.

\acknowledgements

Heartfelt thanks to an anonymous referee who spurred
significant improvements in the manuscript and a deep
mahalo to the queue observers for the Keck-Gemini NIRSPEC
program, Tom Geballe and Marianne Takamiya for the 2001
data. NASA grant NAG5-12996 issued through the Office of
Space Science provides support for this project.

\clearpage

\begin{figure}
\rotatebox{-90}{
\epsscale{0.8}
\plotone{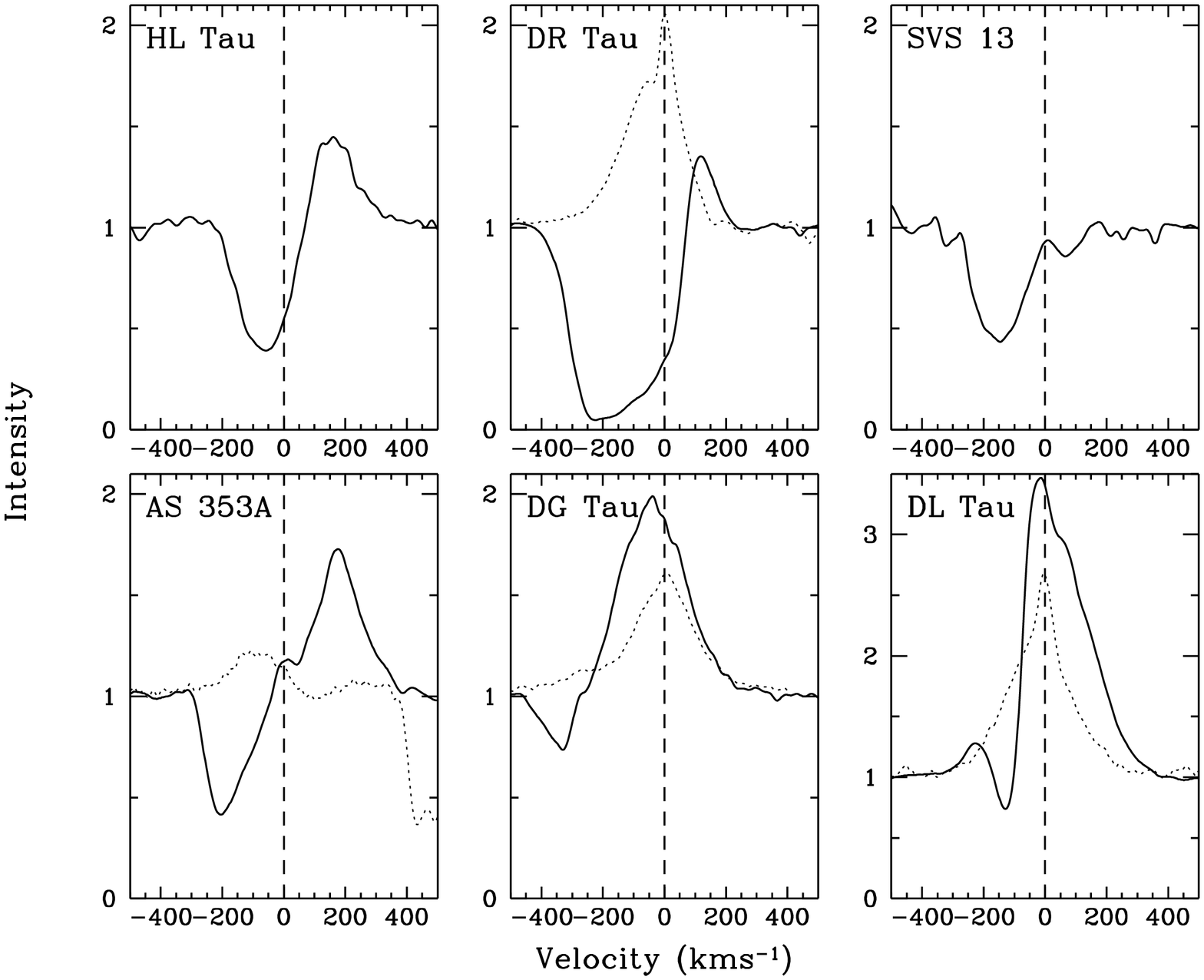}
\epsscale{1.0}}
\caption{Solid lines show He I $\lambda$10830 profiles for 6
low mass young stars with high disk accretion rates. Spectra are
normalized to the continuum intensity and registered at the
velocity of the ambient molecular cloud, known to be within a
few km~s$^{-1}$ of the stellar velocity for Taurus YSO's
\citep{lee86}. For 4 of the stars non-simultaneous He I $\lambda$5876
profiles, normalized to their continuum intensity, are also shown
as dotted lines. }
\label{fig1} \end{figure}

\begin{figure}
\rotatebox{-90}{
\epsscale{0.8}
\plotone{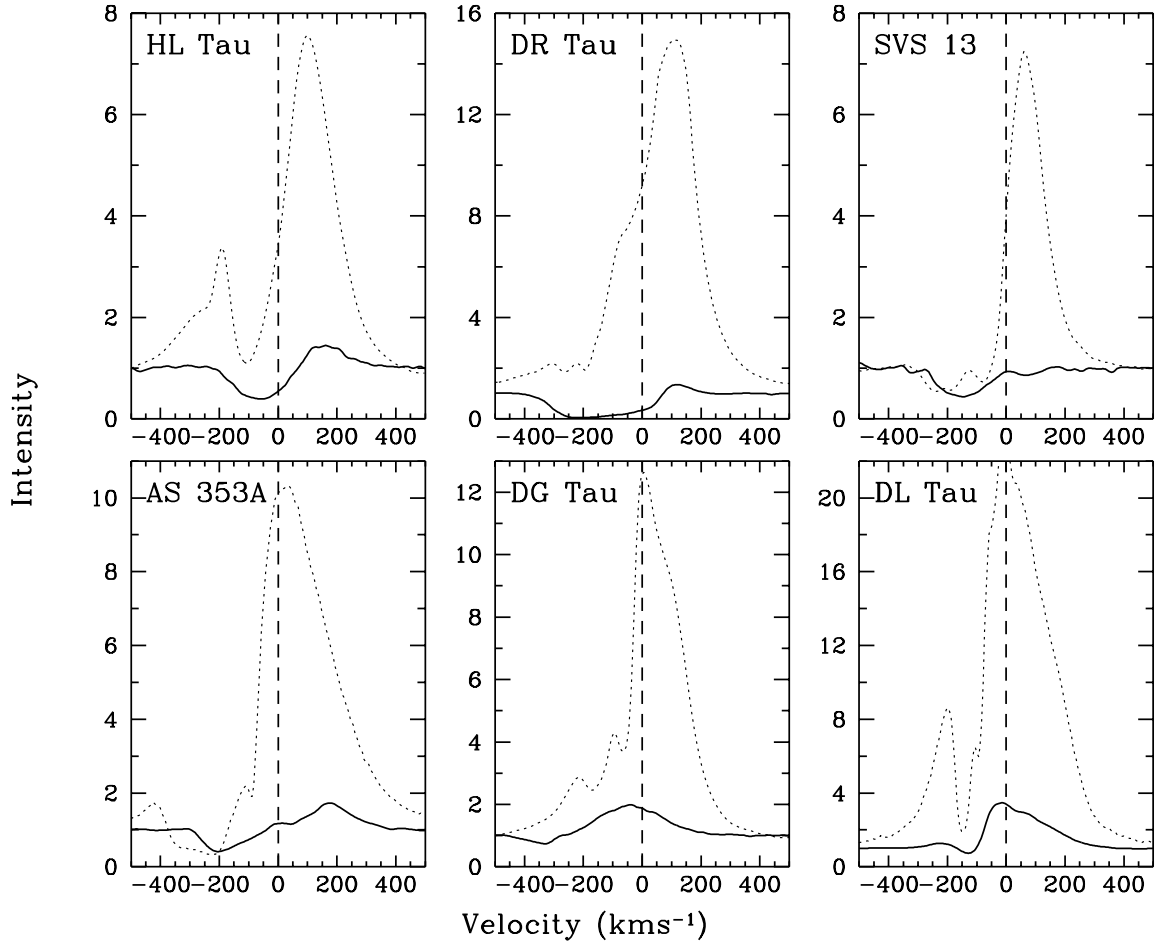}}
\caption{The same 6 He I $\lambda$10830 profiles from Figure 1
(solid lines) are re-scaled for comparison to non-simultaneous
profiles of H$\alpha$ (dotted lines).
At H$\alpha$ the profiles are overwhelmingly in emission while at He I
blueshifted absorption is a far more prominent contributor. }
\label{fig2} \end{figure}

\clearpage

\begin{deluxetable}{lrccccccc}
\tabletypesize{\scriptsize}
\tablecaption{Classical T Tauri Stars \label{tbl-1}}
\tablewidth{0pt}
\tablehead{
\colhead{Star} &
\colhead{$V_{\star}$\tablenotemark{a}} &
\colhead{$-$W$_{\lambda}$ \tablenotemark{b}}   &
\colhead{$+$W$_{\lambda}$ \tablenotemark{c}}   &
\colhead{Net W$_{\lambda}$ \tablenotemark{d}}   &
\colhead {$V_{max}$ \tablenotemark{e}} &
\colhead {$V_{jet}$ \tablenotemark{f}} &
\colhead {$r_Y$ \tablenotemark{g}} &
\colhead{log ${\dot M}_{acc}$\tablenotemark{h}} \\
\colhead{} & \colhead{km~s$^{-1}$}  &
\colhead{\AA} & \colhead{\AA} &  \colhead{\AA} &
\colhead{km~s$^{-1}$} &
\colhead{km~s$^{-1}$} & \colhead{}  &
\colhead{$M_{\odot}yr^{-1}$}\\
}
\startdata
AS353A &-10 &-3.1 & 4.0 &0.9 &  300 & 300 & 2.2& -5.4 \\
DG Tau &+16 &-1.0 & 8.9 &7.9 & 450 & 250 &1.8& -5.7 \\
DL Tau &+16 &-2.5 &19.0 &16.5 & 200 & 200 &1.0& -6.7 \\
DR Tau &+22 &-11.3& 1.1 &-10.2& 400 & 200 &2.6& -5.1 \\
HL Tau &+18 &-4.0 & 2.5 &-1.5& 200 & 180 &1.0& -5.4 \\
SVS13  &+13 &-4.0 & 0  &-4.0 & 250 & 150 &\nodata& -6.2: \\
\enddata


\tablecomments{
$^a$ Heliocentric velocity of ambient molecular gas
used for stellar rest velocity
from Edwards \& Snell 1982 \& Snell \& Edwards 1981;
$^b$ Equivalent width below the continuum;
$^c$ Equivalent width above the contiuum;
$^d$ Net equivalent width;
$^e$ Maximum velocity of He I 10830 blueshifted absorption;
$^f$ Jet radial velocities from HEG, Mundt et al. 1990, \& Davis
et al. 2003;
$^g$ Veiling in the Y band at 1$\mu m$;
$^h$ Mass accretion rates from HEG, Calvet et al .1994,
and estimated for SVS13 by assuming the accretion rate is 10$\times$ the
published mass loss rate from Davis et al. 2001.}

\end{deluxetable}

\end{document}